\documentclass[12pt,a4paper]{article}

\newlength{\dinwidth}
\newlength{\dinmargin}
\usepackage{epsfig}
\usepackage{amsmath}
\usepackage{amssymb}
\usepackage{cite}
\usepackage{setspace}

\def\ie{\textsl{i.e.}}
\def\eg{\textsl{e.g.}}
\newcommand{\comment}[1]  {  }
\def\BE{\begin{equation}}
\def\EE{\end{equation}}
\def\BEA{\begin{eqnarray}}
\def\EEA{\end{eqnarray}}

\newcommand{\pd}[2]{\frac{\partial #1}{\partial #2}}

\newcommand\vx{{\bf x}}

\begin{document}

% paper title
\title{\textbf{On the Achievable Information Rates of\\CDMA Downlink with Trivial Receivers}}

\author{Ori~Shental\footnote{Department of Electrical Engineering-Systems,
Tel-Aviv University, Tel-Aviv 69978, Israel \newline(e-mail: \{shentalo,ajw\}@eng.tau.ac.il).}\,, Ido
Kanter\footnote{Minerva Center and Department of Physics, Bar-Ilan University, Ramat-Gan 52900, Israel
\newline(e-mail: kanter@mail.biu.ac.il).\newline This research was supported by the Israel Science Foundation
(Grants 1232/04, 296/03).\newline\textbf{Submitted to IEEE Transactions on Information Theory.}}\,, and Anthony
J. Weiss$^{\ast}$}
\date{\normalsize{September 10, 2005}}
% make the title area
\maketitle

\vspace{-10mm} \doublespacing
\begin{abstract}
A noisy CDMA downlink channel operating under a strict complexity
constraint on the receiver is introduced. According to this
constraint, detected bits, obtained by performing hard decisions
directly on the channel's matched filter output, must be the same
as the transmitted binary inputs. This channel setting, allowing
the use of the simplest receiver scheme, seems to be worthless,
making reliable communication at any rate impossible. However,
recently this communication paradigm was shown to yield valuable
information rates in the case of a noiseless channel. This finding
calls for the investigation of this attractive
complexity-constrained transmission scheme for the more practical
noisy channel case. By adopting the statistical mechanics notion
of metastable states of the renowned Hopfield model, it is proved
that under a bounded noise assumption such complexity-constrained
CDMA channel gives rise to a non-trivial Shannon-theoretic
capacity, rigorously analyzed and corroborated using finite-size
channel simulations. For unbounded noise the channel's outage
capacity is addressed and specifically described for the popular
additive white Gaussian noise.
\end{abstract}


\newpage
\section{Introduction}
Code-division multiple-access (CDMA) technology serves extensively in wireless communication systems. As such,
revealing its information-theoretic properties has been a fruitful source of ongoing research (\eg,
~\cite{BibDB:VerduShamai,BibDB:Tanaka} and references therein).

As in any problem of reliable (\ie\,errorless) communication via a detrimental channel, typical
information-theoretic investigation of CDMA channels must be carried out under certain resource constraints. For
instance, often an upper bounded transmission power or limited bandwidth are assumed, but usually no
restrictions on complexity are imposed.

However, in the era of pervasive and ubiquitous communications there is an emerging interest in the workings of
a stricter complexity-constrained scenario. According to this scenario, in the CDMA downlink detected bits,
sliced at the output of the user's matched filter, must be the same as the transmitted binary inputs. Such a
transmission constraint implies the appealing use of low-cost trivial receivers. Still, this channel setting
seems at first to be worthless, making reliable communication at any rate impossible.

Recently~\cite{BibDB:ShentalCDMA}, we have computed the Shannon capacity of a \emph{noiseless}
complexity-constrained CDMA channel and found it to yield non-trivial capacity. In some cases, the capacity of
the noiseless complexity-constrained channel was proved to be comparable to the capacity of optimal multi-user
receiver. Nevertheless, formerly there has been no rigorous examination of the information-theoretic
characteristics of the more practical \emph{noisy} CDMA channels under this strict user complexity-constrained
setting.

In this paper, we extend our previous work~\cite{BibDB:ShentalCDMA} and compute the capacity of a noisy CDMA
downlink with trivial mobile receivers. For this purpose, we borrow analysis tools from equilibrium statistical
mechanics, especially the Hopfield model of neural networks~\cite{BibDB:Gardner,BibDB:Singh}. Note that although
the Hopfield model has been utilized in previous works for developing sub-optimal multi-user
detectors~\cite{BibDB:KechriotisEtAl,BibDB:Tanaka2,BibDB:Tanaka3}, this contribution (along
with~\cite{BibDB:ShentalCDMA}) is the first attempt to exploit the metastable states structure of the Hopfield
model for Shannon-theoretic investigation of CDMA and communication channels in general.
%Statement of Value
By borrowing this statistical mechanics notion, valuable achievable information rates are unveiled.

The  paper is organized as follows. Section~\ref{sec_model} introduces the noisy complexity-constrained CDMA
channel model, while section~\ref{sec_capacity} derives rigorously its asymptotic capacity.
Section~\ref{sec_results} provides and discusses the devised analytical capacity curves, being validated via
finite-size simulations. We conclude in section~\ref{sec_conclusion}.

We shall use the following notations. The operator $f'(\cdot)$ denotes a derivative of $f(\cdot)$ with respect
to (w.r.t.) its argument, while $<\cdot>_{\vx}$ denotes the average w.r.t. $\vx$, and $\delta(\cdot)$ is the
Dirac delta function. The symbols $j\triangleq\sqrt{-1}$, $\omega$ is the angular frequency of the Fourier
transform, while $\sum_{\vx}$ and $\sum_{i\neq k}$ correspond to a sum over all the possible values of $\vx$ and
a non-overlapping summation, respectively. Finally, an error function is defined by
$Q(x)\triangleq1/\sqrt{2\pi}\int_{x}^{\infty}dy\exp{(-y^{2}/2)}$.

\section{Channel Model}\label{sec_model}
\begin{figure}[htb]\centerline{\includegraphics[width=1\textwidth]{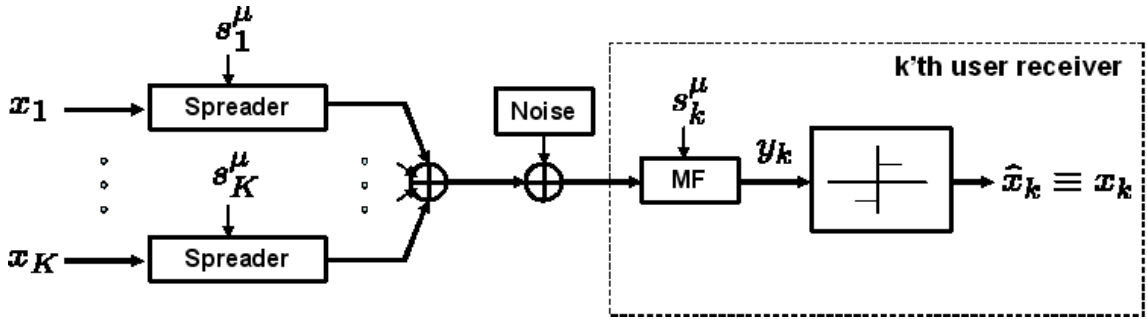}}\vspace*{-170mm}
\caption{Discrete-time schematic of the complexity-constrained CDMA downlink.}\label{fig_model}
\end{figure}

Consider a noisy synchronous CDMA downlink (depicted in
Fig.~\ref{fig_model}) accessing $K$ active users via the mutual
channel in order to transmit their designated (coded) information
binary symbols, $x_{k}=\pm1$, where $k=1,\ldots,K$, with equal
power $P$. Each transmission to a user is assigned with a binary
signature sequence (spreading code) of $N$ chips,
$s_{k}^{\mu}=\pm1$, $\mu=1\ldots N$.

Assuming a random spreading model, the binary chips are independently equiprobably chosen, and the deterministic
chip waveform has unit energy. The cross-correlation between users' transmissions is
\mbox{$\rho_{ki}\triangleq1/N\sum_{\mu}s_{k}^{\mu}s_{i}^{\mu}$}.

The received signal is passed through the user's matched filter (MF). Thus, the overall downlink channel is
described by \BE\label{eq_channel}
    y_{k}=\sqrt{P}x_{k}+\sqrt{P}\sum_{i\neq k}\rho_{ki}x_{i}+n_{k},
\EE where the $k$'th user matched filter output, $y_{k}$, is the
designated bit, $x_{k}$, corrupted by interference and noise
terms. The interference term is composed of a summation over
cross-correlation scaled versions of all other users' bits. The
set of all cross-correlations $\{\rho_{ki}\}$ is hereinafter
denoted by $\rho$. The noise term is assumed to be independent and
identically distributed (i.i.d.) and symmetrically bounded, \ie\,
\mbox{$-\kappa\sqrt{P}<n_{k}<\kappa\sqrt{P}$}, where the threshold
$\kappa$ is a known non-negative constant. In the following
asymptotic analysis, we assume that $K\rightarrow\infty$, yet the
system load factor $\beta\triangleq K/N\triangleq\alpha^{-1}$ is
kept constant, and that the information rate is the same for all
users, \ie\, $R_{k}=R$.

\vspace{10mm}
\vspace{10mm}

We want to convey information reliably through the
channel~(\ref{eq_channel}) under a low-complexity constraint on
the user receiver. According to this constraint, detected bits,
$\hat{x}_{k}$, obtained by performing hard decisions directly on
the single-user matched filter output samples, must be the same as
the transmitted bits. Explicitly, \BE\label{eq_constraint}
    x_{k}\equiv\hat{x}_{k}=\emph{{\textrm{Sign}}}(y_{k}),
\EE where $\textrm{Sign}(\cdot)$ is the trivial sign function.

For any $x_{k}$ and $y_{k}$ which maintain the constraint~(\ref{eq_constraint}), there is a certain
\emph{positive} scalar $\lambda_{k}$ for which \BE
    x_{k}y_{k}=\lambda_{k}>0.
\EE Moreover, the bounded noise assumption yields that \BE
    x_{k}\big(x_{k}+\sum_{i\neq k}\rho_{ki}x_{i})\big)>\kappa,
\EE or alternatively \BE\label{eq_constraint2}
    \sum_{i\neq k}\rho_{ki}x_{k}x_{i}>\kappa-1.
\EE

It is important to notice that due to the trivial receiver scheme adopted, and especially its non-linear sign
operation, the Shannon capacity becomes invariant in the exact noise probability distribution function, but it
only depends on the noise upper and lower values $\pm\kappa\sqrt{P}$.\footnote{In the case of unbounded noise,
e.g. additive white Gaussian noise (AWGN), the following analysis still holds, but the term of Shannon capacity
should be replaced with the term of outage capacity. This issue is addressed in section~\ref{sec_results}.} This
explains why the bounded noise definition, regardless of its exact probability distribution function form,
suffices.

Under the constraints outlined above it is clear that not all combinations of input symbols will result in
errorless communication. Thus, the capacity of the channel can be obtained by evaluating the number of codewords
that ensure errorless detection. The codewords constraint~(\ref{eq_constraint2}) is analogous to the constraint
on the single-neuron (or spin) flip metastable states of the Hopfield model. In the following section we prove
that this complexity-constrained CDMA channel setting yields non-trivial capacity.

\section{Asymptotic Capacity}\label{sec_capacity}
A binary codeword $\vx^{c}\triangleq\{x_{1}^{c},\ldots,x_{K}^{c}\}$, composed of all $K$ users' bits at a given
channel use, for which the channel constraints~(\ref{eq_constraint2}) hold, satisfies the condition
\BE\label{eq_stable}
    \int_{\kappa-1}^{\infty}\prod_{k=1}^{K}d\lambda_{k}\delta(\sum_{i\neq k}\rho_{ki}x_{i}-\lambda_{k}x_{k})
    =1. \EE Condition~(\ref{eq_stable}) can be
reformulated as \BEA
    &&\alpha\int_{\kappa-1}^{\infty}\prod_{k=1}^{K}d\lambda_{k}\delta(\sum_{i\neq
    k}\alpha\rho_{ki}x_{i}-\alpha\lambda_{k}x_{k})\\&=&
    \int_{\alpha(\kappa-1)}^{\infty}\prod_{k=1}^{K}d\lambda_{k}\delta(\sum_{i\neq k}\alpha\rho_{ki}x_{i}-\lambda_{k}x_{k})=1\nonumber.
\EEA

Let the random variable $\mathbb{N}(K,\beta,\rho,\kappa)$ denote the number of codewords, \ie
\BE\label{eq_number0}
    \mathbb{N}(K,\beta,\rho,\kappa)\triangleq\\\int_{\alpha(\kappa-1)}^{\infty}\prod_{k=1}^{K}d\lambda_{k}\delta(\sum_{i\neq k}\alpha\rho_{ki}x_{i}-\lambda_{k}x_{k}).
\EE Assuming equal user information rates, the corresponding asymptotic capacity of the channel is
defined~\cite{BibDB:BookCover}, in bit information units, as \BE\label{eq_capacity}
    C_{\infty}(\beta,\kappa)\triangleq\lim_{K\rightarrow\infty}\frac{\log_{2}{\mathbb{N}(K,\beta,\rho,\kappa)}}{K}.
\EE Assuming self-averaging property~\cite{BibDB:Tanaka,BibDB:BookEllis}, in the large-system limit
$K\rightarrow\infty$ the number of successful codewords $\mathbb{N}(K,\beta,\rho,\kappa)$ is equal to its
expectation w.r.t. the distribution of $\rho$, \ie \BEA\label{eq_number1}
    &&\lim_{K\rightarrow\infty}\mathbb{N}(K,\beta,\rho,\kappa)=\mathcal{N}(\beta,\kappa)\nonumber\\&=&
    \lim_{K\rightarrow\infty}\int_{\alpha(\kappa-1)}^{\infty}\prod_{k}d\lambda_{k}
    \sum_{\vx}\Bigg<\prod_{k}\delta\Big(\sum_{i\neq k}\alpha\rho_{ki}x_{i}-\lambda_{k}x_{k}\Big)\Bigg>_{\rho},
\EEA where $\mathcal{N}(\beta,\kappa)$ denotes the average.

Representing the delta function by the inverse Fourier transform of an exponent, expression~(\ref{eq_number1})
can be rewritten as \BEA\label{eq_number2}
    \mathcal{N}(\beta,\kappa)&=&\lim_{K\rightarrow\infty}\int_{\alpha(\kappa-1)}^{\infty}\prod_{k}d\lambda_{k}
    \frac{1}{(2\pi)^{K}}\int_{-\infty}^{\infty}\prod_{k}d\omega_{k}\nonumber\\
    &\times&\sum_{\vx}\exp{\Big(j\sum_{k}{\omega_{k}\lambda_{k}x_{k}}\Big)}
    \Bigg<\exp{\Big(-j\sum_{i\neq k}\alpha\rho_{ki}x_{i}\omega_{k}\Big)}\Bigg>_{\rho}.
\EEA Substituting $x_{k}\omega_{k}$ for $\omega_{k}$, we find \BEA\label{eq_number3}
    \mathcal{N}(\beta,\kappa)&=&\lim_{K\rightarrow\infty}\int_{\alpha(\kappa-1)}^{\infty}\prod_{k}d\lambda_{k}
    \frac{1}{(2\pi)^{K}}\int_{-\infty}^{\infty}\prod_{k}d\omega_{k}\nonumber\\
    &\times&\sum_{\vx}\exp{\Big(j\sum_{k}{\omega_{k}\lambda_{k}}\Big)}
    \cdot\mathbb{E},
\EEA where \BEA\label{eq_expectation}
    \mathbb{E}&\triangleq&\Bigg<\exp{\Big(-j\sum_{i\neq k}\alpha\rho_{ki}x_{i}x_{k}\omega_{k}\Big)}\Bigg>_{\rho}\nonumber\\
    &=&\Bigg<\exp{\Big(-j\sum_{i\neq k}\frac{1}{K}\sum_{\mu=1}^{N}s_{k}^{\mu}s_{i}^{\mu}x_{i}x_{k}\omega_{k}\Big)}\Bigg>_{\rho}
.
\EEA The expectation $\mathbb{E}$ can be also written as \BEA\label{eq_expectation11}
    \mathbb{E}&=&\exp{(j\alpha\sum_{k}\omega_{k})}\Bigg<\exp{\Big(-\frac{j}{K}\sum_{\mu}(\sum_{k}s_{k}^{\mu}x_{k}\omega_{k})(\sum_{k}s_{k}^{\mu}x_{k})\Big)}\Bigg>_{\rho}.
\EEA

Using the transformation~\cite{BibDB:BruceEtAl} \BEA
    \exp{(-jA_{\mu}B_{\mu}/K)}&=&\int_{-\infty}^{\infty}\frac{da_{\mu}}{(2\pi/K)^{1/2}}
    \int_{-\infty}^{\infty}\frac{db_{\mu}}{(2\pi/K)^{1/2}}\\&\times&
    \exp{\Big(j\frac{K}{2}(a_{\mu}^{2}-b_{\mu}^{2})-\frac{j}{\sqrt{2}}A_{\mu}(a_{\mu}+b_{\mu})-\frac{j}{\sqrt{2}}B_{\mu}(a_{\mu}-b_{\mu})\Big)}\nonumber,
\EEA expression~(\ref{eq_expectation11}) becomes (here, and hereafter, logarithms are taken to base
$e$)\BEA\label{eq_expectation2}
    \mathbb{E}&=&\exp{(j\alpha\sum_{k}\omega_{k})}\int_{-\infty}^{\infty}\prod_{\mu}\frac{da_{\mu}}{(2\pi/K)^{1/2}}
    \int_{-\infty}^{\infty}\prod_{\mu}\frac{db_{\mu}}{(2\pi/K)^{1/2}}\nonumber\\&\times&\exp{\Big(j\frac{K}{2}\sum_{\mu}(a_{\mu}^{2}-b_{\mu}^{2})
    +\sum_{k,\mu}\log\big(\cos(c_{k,\mu})\big)\Big)},
\EEA where \BE
    c_{k,\mu}\triangleq\frac{1}{\sqrt{2}}\big(\omega_{k}(a_{\mu}+b_{\mu})+(a_{\mu}-b_{\mu})\big).
\EE

Since $\sum_{k}s_{k}^{\mu}x_{k}$ in~(\ref{eq_expectation11}) is $\mathcal{O}(\sqrt{K})$ for a vast majority of
codewords, for the expectation $\mathbb{E}$ to be finite, $a_{\mu}$ and $b_{\mu}$ must be
$\mathcal{O}(1/\sqrt{K})$. Hence, expanding the $\log\big(\cos(\cdot)\big)$ term in
exponent~(\ref{eq_expectation2}) and neglecting terms of order $1/K$ and higher, we get
\BEA\label{eq_expectation3}
    \mathbb{E}&=&\exp{(j\alpha\sum_{k}\omega_{k})}\int_{-\infty}^{\infty}\prod_{\mu}\frac{da_{\mu}}{(2\pi/K)^{1/2}}
    \int_{-\infty}^{\infty}\prod_{\mu}\frac{db_{\mu}}{(2\pi/K)^{1/2}}\nonumber\\&\times&\exp{\Big(j\frac{K}{2}\sum_{\mu}(a_{\mu}^{2}-b_{\mu}^{2})
    -\frac{1}{4}\sum_{k,\mu}\hat{c}_{k,\mu}\Big)}, \EEA
where \BE
    \hat{c}_{k,\mu}\triangleq\big(\omega_{k}^{2}(a_{\mu}+b_{\mu})^{2}+2\omega_{k}(a_{\mu}^{2}-b_{\mu}^{2})+(a_{\mu}-b_{\mu})^{2}\big).
\EE

Now, the solution of the $K$-dimensional integral~(\ref{eq_expectation3}) of the expectation $\mathbb{E}$ is
performed using the following mathematical recipe: New variables are introduced\BEA
    a&\triangleq&\frac{1}{2\alpha}\sum_{\mu}(a_{\mu}+b_{\mu})^{2}\label{eq_a},\\
    b&\triangleq&\frac{j}{2\alpha}\sum_{\mu}(a_{\mu}^{2}-b_{\mu}^{2})+1\label{eq_b}.
\EEA Equations~(\ref{eq_a}) and~(\ref{eq_b}) can be reformulated via the integral representation of a delta
function using the corresponding angular frequencies $A$ and $B$, respectively, \BEA
    \int_{-\infty}^{\infty}\frac{da\,dA}{2\pi/K\alpha}\exp{\big(jKA(\alpha
    a-\sum_{\mu}\frac{(a_{\mu}+b_{\mu})^{2}}{2})\big)}&=&1,\label{eq_aF}\\
    \int_{-\infty}^{\infty}\frac{db\,dB}{2\pi/K\alpha}\exp{\big(jKB(\alpha
    b-j\sum_{\mu}\frac{(a_{\mu}^{2}-b_{\mu}^{2})}{2}-\alpha)\big)}&=&1.\label{eq_bF} \EEA

Substituting these (unity) integrals into the expectation expression~(\ref{eq_expectation3}) and rewriting it
using $a$ and $b$, the integrations over $a_{\mu}$ and $b_{\mu}$ are decoupled and can be performed easily.
Next, for the asymptotics $K\rightarrow\infty$, the integration over the frequencies $A$ and $B$ can be
performed algebraically by the saddle-point method~\cite{BibDB:BookEllis}.

According to this method, the main contribution to the integral comes from values of $A$ and $B$ in the vicinity
of the maximum of the exponent's argument. Finally, the $\mathbb{E}$ term boils down to
\BEA\label{eq_expectation_final}
    \mathbb{E}&=&\int_{-\infty}^{\infty}\frac{da\
    db}{4\pi/K\alpha}\exp{\big(-\frac{1}{2}\alpha a\sum_{k}\omega_{k}^{2}+j\alpha b\sum_{k}\omega_{k}\big)}\nonumber\\
    &\times&\exp{\big(K\alpha(b-\frac{1}{2}+\frac{(1-b)^{2}}{2a}+\frac{1}{2}\log{a})\big)}.
\EEA

Substituting the expectation term~(\ref{eq_expectation_final}) back in~(\ref{eq_number3}), the integrand in the
latter becomes independent of $\vx$, and therefore the $\sum_{\vx}$ can be substituted by multiplying with the
scalar $2^K$. Hence, \BEA
    \mathcal{N}(\beta,\kappa)&=&\lim_{K\rightarrow\infty}\int_{\alpha(\kappa-1)}^{\infty}\prod_{k}d\lambda_{k}\frac{1}{\pi^{K}}
    \int_{-\infty}^{\infty}\prod_{k}d\omega_{k}\exp{\Big(j\sum_{k}{\omega_{k}\lambda_{k}}\Big)}
    \nonumber\\&\times&\int_{-\infty}^{\infty}\frac{da\
    db}{4\pi/K\alpha}\exp{\Big(K\alpha\big(b-\frac{1}{2}+\frac{(1-b)^{2}}{2a}+\frac{1}{2}\log{a}\big)\Big)}
    \nonumber\\&\times&\exp{\Big(-\frac{1}{2}\alpha a\sum_{k}\omega_{k}^{2}+j\alpha b\sum_{k}\omega_{k}\Big)},
\EEA where the resulting $\omega$ dependent integrand is a Gaussian function. Thus, performing Gaussian
integration and exploiting the symmetry in the $K$-dimensional space, we get \BEA\label{eq_number4}
    \mathcal{N}(\beta,\kappa)&=&\lim_{K\rightarrow\infty}\frac{1}{\pi^{K}}
    \int_{-\infty}^{\infty}\frac{da\
    db}{4\pi/K\alpha}\exp{\Big(K\alpha\big(b-\frac{1}{2}+\frac{(1-b)^{2}}{2a}+\frac{1}{2}\log{a}\big)\Big)}
    \nonumber\\&\times&\exp{\bigg(K\log{\Big(\sqrt{\frac{2\pi}{\alpha a}}\int_{\alpha(\kappa-1)}^{\infty}d\lambda\exp{\big(-\frac{(\alpha b+\lambda)^{2}}{2\alpha
    a}\big)}\Big)}\bigg)}.\EEA
Using the rescaling $(\alpha b+\lambda)/\sqrt{\alpha a}\rightarrow \lambda$, the integral~(\ref{eq_number4})
becomes \BE\label{eq_integral}
    \mathcal{N}(\beta,\kappa)=\lim_{K\rightarrow\infty}\int_{-\infty}^{\infty}\frac{da\
    db}{4\pi/K\alpha}\exp{\big(Kg(a,b,\beta,\kappa)\big)},
\EE where the function $g(a,b,\beta,\kappa)$ is defined by \BEA
    g(a,b,\beta,\kappa)\triangleq\frac{1}{\beta}\big(b-\frac{1}{2}+\frac{(1-b)^{2}}{2a}+\frac{1}{2}\log{a}\big)+\log\big(2Q(t)\big),
\EEA with an auxiliary variable \BE
    t\triangleq\frac{\sqrt{\alpha}(b+\kappa-1)}{\sqrt{a}}.
\EE

Again, for $K\rightarrow\infty$, the double integral in~(\ref{eq_integral}) can be evaluated by the saddle-point
method. Hence, we find\footnote{The exponent pre-factor in~(\ref{eq_finalNumber}) is not required for computing
the asymptotic capacity, and therefore it is omitted.} \BE\label{eq_finalNumber}
    \mathcal{N}(\beta,\kappa)\propto\lim_{K\rightarrow\infty}\exp{\big(Kg(a^{\ast},b^{\ast},\beta,\kappa)\big)},
\EE where $a^{\ast}$ and $b^{\ast}$ are found by the saddle-point conditions, which yield the following
equations \BEA
    \pd{g(a,b,\beta,\kappa)}{a}&=&\beta^{-1}\big(\frac{(1-b)^{2}}{a}-1\big)+t\frac{Q'(t)}{Q(t)}=0,\\
    \pd{g(a,b,\beta,\kappa)}{b}&=&\beta^{-1}\big(1-\frac{1-b}{a}\big)+\frac{1}{\sqrt{a\beta}}\frac{Q'(t)}{Q(t)}=0.
\EEA

One then finds that this set of equations is satisfied by \BEA
    b^{\ast}&=&0,\\
    a^{\ast}&=&\beta^{-1}/\big({\beta^{-1}+\frac{1}{\sqrt{a^{\ast}\beta}}\frac{Q'(t^{\ast})}{Q(t^{\ast})}}\big),\label{eq_fix_a}
\EEA where \BE
    t^{\ast}\triangleq(\kappa-1)/{\sqrt{a^{\ast}\beta}}.
\EE This saddle-point condition's fixed-point $a^{\ast}$ can be found iteratively, and it always converges in
the examined model~\cite{BibDB:Singh}.

Finally, substituting~(\ref{eq_finalNumber}) into (\ref{eq_capacity}) the asymptotic capacity, in nat per symbol
per user, is now easily obtained \BEA\label{eq_C}
    C_{\infty}(\beta,\kappa)=g(a^{\ast},b^{\ast},\beta,\kappa)=\log\big(2Q(t^{\ast})\big)+\frac{1}{\beta}\big(\frac{1}{2a^{\ast}}+\frac{1}{2}\log{a^{\ast}}-\frac{1}{2}\big),
\EEA which forms our pivotal result. In section~\ref{sec_results}
we further discuss the theoretical results and compare them with
computer simulations of the complexity-constrained CDMA channel.

\section{Results and Discussion}\label{sec_results}
\begin{figure}[htb!]
\centerline{\includegraphics[width=1\textwidth]{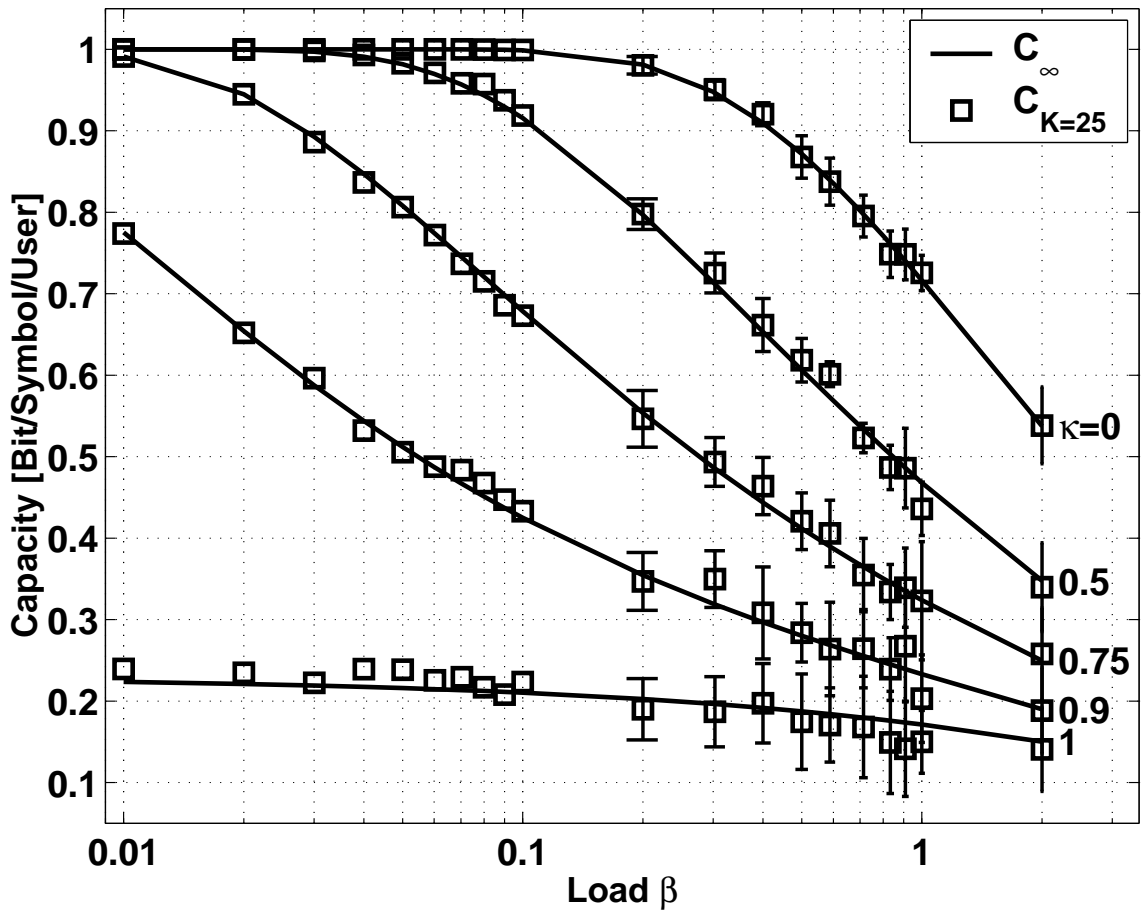}}\caption{Asymptotic
capacity $C_{\infty}$ (solid line), in terms of bit/symbol/user, as a function of load $\beta$ in various noise
levels $\kappa=0,0.5,0.75,0.9,1$. Also drawn is the finite-size simulation-averaged capacity $C_{K}$ for $K=25$
(empty squares). Vertical bars stand for standard deviation in simulation results.}\label{fig_capacity}
\end{figure}

Fig.~\ref{fig_capacity} displays the asymptotic capacity
$C_{\infty}$~(\ref{eq_C}), obtained by solving iteratively the
saddle-point condition~(\ref{eq_fix_a}), as a function of the load
$\beta$ in various noise levels. Interestingly, in noiseless
channel ($\kappa=0$) for small $\beta\lesssim0.1$ values the
trivial 1 bit upper bound (of an optimal receiver, \ie\,matrix
inversion) is practically achieved by this simple hard decision
operation.

Furthermore, for higher system load such a complexity-constrained CDMA setting still yields substantial
achievable information rates. Note, in passing, that for heavily overloaded system
(\ie\,$\beta\rightarrow\infty$) the capacity curve decay of the noiseless case coincides with Hopfield model's
capacity (see~\cite[eq.~(12)]{BibDB:Gardner} for an analytical approximation of this capacity decay to zero.)
Even in the presence of noise non-negligible rates are obtained for the examined noise levels (up to $\kappa=1$)
in a wide range of load values.

In order to validate the analytically derived asymptotic capacity
$C_{\infty}(\beta,\kappa)$, we evaluated the capacity
$C_{K}(\beta,\kappa)$ of a noisy CDMA downlink channel with large,
yet finite number of users $K$, using exhaustive search
simulations. The number of successful binary codewords,
maintaining the channel constraints~(\ref{eq_constraint2}), was
obtained by examining all $2^K$ possible codewords. The average
logarithm of the counted number, normalized by the number of users
$K$, gives the capacity $C_{K}$.

Fig.~\ref{fig_capacity} presents the capacity obtained by simulations for $K=25$. As can be seen, the empirical
capacity for finite $K$ deviates only slightly from the analytically obtained asymptotic capacity. Due to
finite-size effects these slight deviations from theoretical results grow with the decrease in the capacity.
These results substantiate the analysis of the complexity-constrained CDMA channel.

\begin{figure}[htb]
\centerline{\includegraphics[width=1\textwidth]{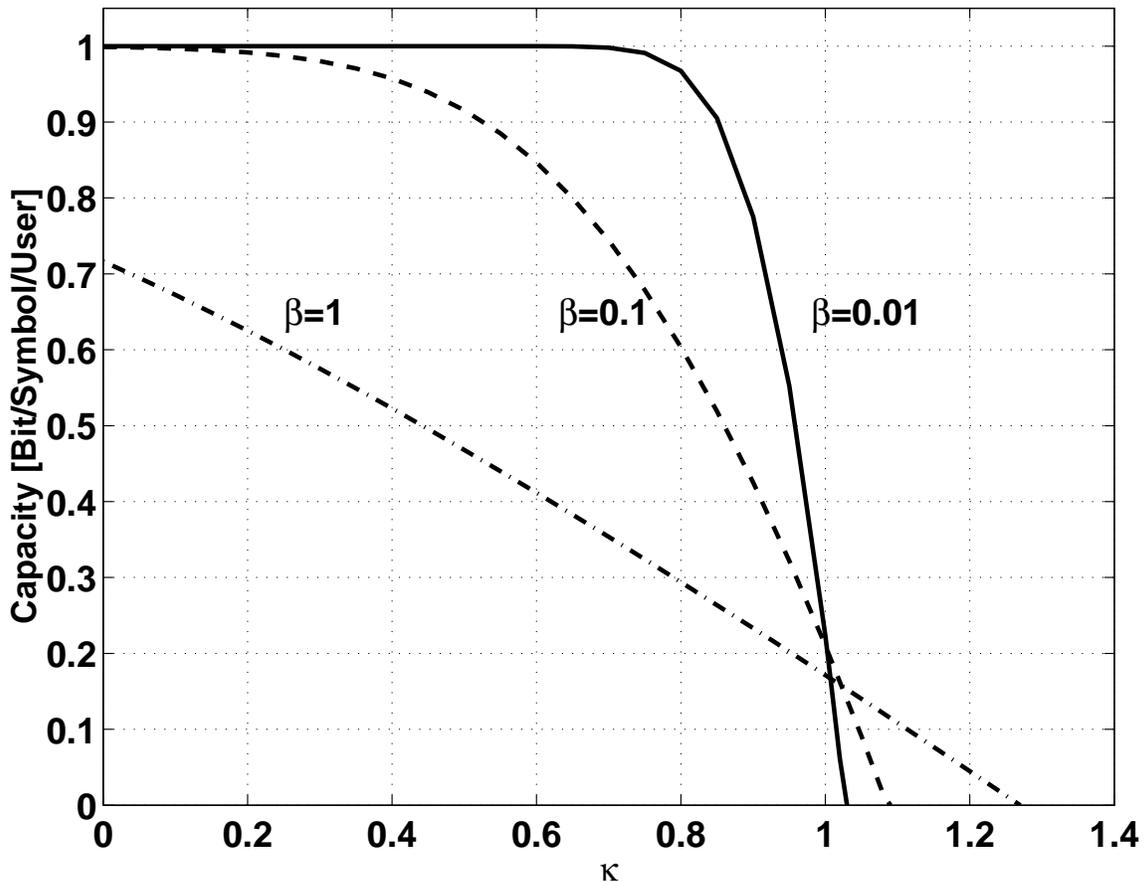}}\caption{\label{fig_capacity2}
Asymptotic capacity $C_{\infty}$, in terms of bit/symbol/user, as a function of noise levels $\kappa$ for a
fixed load \mbox{$\beta=0.01\,\textrm{(solid line)},0.1\,\textrm{(dashed)},1\,\textrm{(dashdotted)}$}.}
\end{figure}

The devised capacity is also drawn in a reciprocal manner.
Fig.~\ref{fig_capacity2} displays $C_{\infty}$ as a function of
the noise threshold $\kappa$, this time for a fixed load $\beta$.
The capacity decreases monotonically as a function of $\kappa$
from asymptotically $1$ bit down to zero capacity.

Three typical regimes can be readily observed: As may be expected, for noise thresholds $\kappa\lesssim1$, an
increase in system load $\beta$ directly results in lower capacity. On the other hand, for thresholds
$\kappa\gtrsim1$, as the system becomes more loaded, the maximum achievable rate increases, and the interfering
users play a \emph{constructive}, rather than destructive role.

This fascinating phenomenon can be explained by the fact that when
the noise becomes more dominant (\ie\,at the order of information
power $P$), a certain user's designated information bit can not
exceed the transmission constraint by itself and needs the
assistance of the "interference" term (organized properly) in
order to deliver its own information reliably.

For \mbox{$\beta=0.01$, $0.1$} and $1$ zero capacity is found to
be inevitable starting from noise thresholds $\approx1.05$, $1.09$
and $1.27$, respectively, for which reliable communication in this
complexity-constrained setting becomes infeasible. The transition
between these two regimes occurs at the vicinity of $\kappa=1$,
for which the capacity is approximately $0.2$ bit, regardless of
the examined system load (as can be seen more clearly from
Fig.~\ref{fig_capacity}.)

\subsection{AWGN and Outage Capacity}
Evidently, for unbounded noise distribution, like the popular additive white Gaussian noise (AWGN), Shannon
capacity, under the examined complexity constraint, is zero. However, the analysis is still useful for obtaining
the outage capacity instead of the Shannon capacity.

\begin{figure}[htb]
\centerline{\includegraphics[width=1\textwidth]{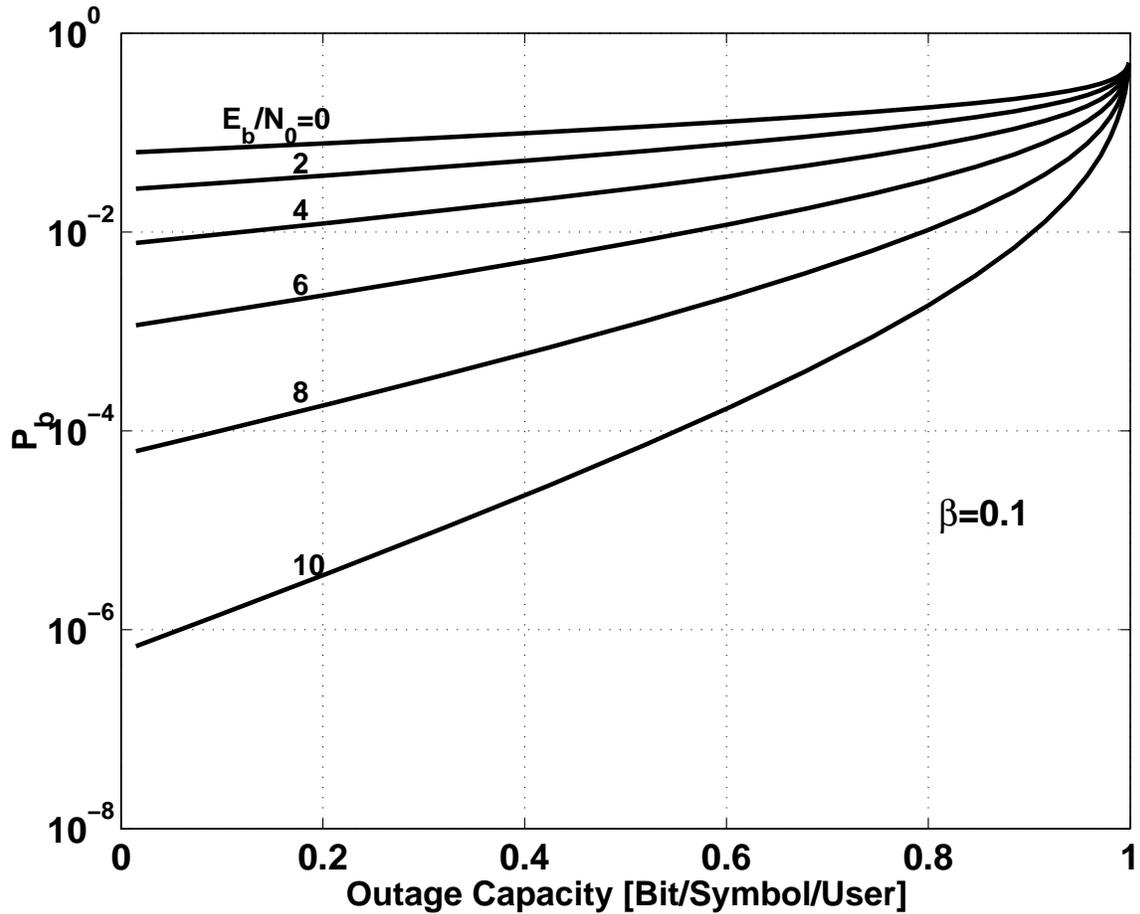}}
\caption{\label{fig_outageCapacity}
    Outage capacity for AWGN: bit error rate per user $P_{b}$ as a function of the rate, in terms of bit/symbol/user,
    for various $E_{b}/N_{0}$ levels ($\beta=0.1$).}
\end{figure}

Fig.~\ref{fig_outageCapacity} presents the bit error rate (BER)
per user $P_{b}$ as a function of the corresponding outage
capacity, in terms of bit/symbol/user, for different
signal-to-noise ratios in the case of AWGN. The BER is evaluated
by computing the probability $\Pr(n_{k}>\kappa\sqrt{P})$, and then
it is linked to a certain achievable information rate via the
analytically derived capacity-threshold dependency (\eg\,the
curves in Fig.~\ref{fig_capacity2}). It can be seen that
reasonable information rates can be achieved.

For instance, for $E_{b}/N_{0}=10$dB a BER of 0.001 (which is the
BER typically required for voice traffic) can be reached with a
rate of 0.75 bit. For comparison, without any complexity
constraint the Shannon capacity for binary-input AWGN CDMA is
asymptotically 1 bit~\cite{BibDB:Tanaka}. However, in order to
approach this capacity an optimal multiuser receiver with
intractable complexity of $\mathcal{O}(2^{K})$ is required, along
with a sophisticated decoding mechanism, while at the cost of 0.25
bit the proposed trivial receiver will do (for voice traffic).

\section{Conclusion}\label{sec_conclusion}
We evaluated the asymptotic capacity of a noisy CDMA downlink channel model requiring only minimal signal
processing at the receiver, thus suitable for networks with low-complexity mobile equipment. Interestingly, we
found a range of non-trivial achievable rates.

According to these findings, at a given channel use a fraction of the users, equal to $C_{\infty}$ (in bit), can
receive its designated information with rate $1$, while the transmissions to the rest of the users ensure
reliable communication. Determining these redundant transmissions in a diagrammatic manner (rather than via
brute-force enumeration, which becomes infeasible for large $K$) remains an interesting open research question.
Also, the method used can be employed in the investigation of other (non-CDMA) noisy complexity-constrained
channels.

\end{document}